\documentclass[aps,prd,amsmath,floats,floatfix, twocolumn,
superscriptaddress,nofootinbib,showpacs]{revtex4-1}

\usepackage[colorlinks, pdfborder={0 0 0}, plainpages=false]{hyperref}
\usepackage{graphicx}
\usepackage{xspace}
\usepackage[usenames,dvipsnames]{color}
\usepackage{amssymb}
\usepackage[normalem]{ulem} 
\usepackage{microtype}

\newcommand{\specialcell}[2][c]{%
  \begin{tabular}[#1]{@{}c@{}}#2\end{tabular}}

\newcommand{\Caltech}{\affiliation{Theoretical Astrophysics 350-17,
    California Institute of Technology, Pasadena, CA 91125, USA}}
\newcommand{\Cornell}{\affiliation{Cornell Center for Astrophysics and
    Planetary Science, Cornell University, Ithaca, New York 14853,
    USA}} \newcommand{\CITA}{\affiliation{Canadian Institute for
    Theoretical Astrophysics, 60 St.~George Street, University of
    Toronto, Toronto, ON M5S 3H8, Canada}} %
\newcommand{\GWPAC}{\affiliation{Gravitational Wave Physics and
    Astronomy Center, California State University Fullerton,
    Fullerton, California 92834, USA}} %
\newcommand{\AEI}{\affiliation{Max Planck Institute for Gravitational
    Physics (Albert Einstein Institute), Am M\"{u}hlenberg 1, 14476
    Potsdam-Golm, Germany}}
\newcommand{\Ohio}{\affiliation{Department of Physics
  and Astronomy, Ohio University, Athens, Ohio 45701, USA}}

\begin{document}

\title{Measuring the properties of nearly extremal black holes with gravitational waves}

\author{Katerina Chatziioannou} \CITA
\author{Geoffrey Lovelace} \GWPAC
\author{Michael Boyle} \Cornell
\author{Matthew Giesler} \Caltech
\author{Daniel~A.~Hemberger} \Caltech
\author{Reza Katebi} \GWPAC\Ohio
\author{Lawrence~E.~Kidder} \Cornell
\author{Harald~P.~Pfeiffer} \CITA\AEI
\author{Mark~A.~Scheel} \Caltech
\author{B\'{e}la~Szil\'{a}gyi} \Caltech

\date{\today}

\begin{abstract}
  Characterizing the properties of black holes is one of the most important science
  objectives for gravitational-wave observations. Astrophysical
  evidence suggests that black holes that are nearly extremal
  (i.e. spins near the theoretical upper limit) might exist and, thus,
  might be among the merging black holes observed with gravitational
  waves. In this paper, we explore how well current gravitational wave
  parameter estimation methods can measure the spins of rapidly
  spinning black holes in binaries. We simulate gravitational-wave
  signals using numerical-relativity waveforms for nearly-extremal,
  merging black holes. For simplicity, we confine our attention to
  binaries with spins parallel or antiparallel with the orbital
  angular momentum. We find that recovering the holes' nearly extremal
  spins is challenging. When the spins are nearly extremal
    and parallel to each other, the resulting parameter estimates do
    recover spins that are large, though the recovered spin magnitudes
    are still significantly smaller than the true spin magnitudes.
    When the spins are nearly extremal and antiparallel to each other,
    the resulting parameter estimates recover the small effective spin
     but incorrectly estimate the individual spins as nearly
    zero.  We study the effect of spin priors and argue that a
  commonly used prior (uniform in spin magnitude and direction)
  hinders unbiased recovery of large black-hole spins.
\end{abstract}


\maketitle

\section{Introduction}

Beginning with the first discovery of gravitational waves (GWs)
passing through Earth in 2015, to date the Laser Interferometer
Gravitational-Wave Observatory (LIGO)~\cite{TheLIGOScientific:2014jea}
and Virgo~\cite{TheVirgo:2014hva} have announced five detections of GWs from merging
binary black holes
(BBH)~\cite{2016PhRvL.116f1102A,2016PhRvL.116x1103A,2017PhRvL.118v1101A,Abbott:2017oio,Abbott:2017gyy}. As
LIGO and Virgo approach their design sensitivity, they are expected to
detect hundreds of merging BH
binaries~\cite{2010CQGra..27q3001A,2017PhRvL.118v1101A}.

One important objective of the GW observations is the measurement of the masses and spins of the merging BHs. This is interesting in its own right, but accurate characterization of the systems' properties is also crucial for
astrophysical inference. The masses and the spins of the binary
components can reveal information about the way these binaries
were formed and about the properties of the BH
progenitors. While most formation scenarios predict similar mass
distributions for merging
BHs~\cite{Belczynski:2016obo,Rodriguez:2016avt,Stevenson:2017tfq}, it
has been suggested that spin measurements might be able to offer information
about different formation
channels and the BH progenitor properties, e.g.~\cite{Rodriguez:2016vmx,Kushnir:2016zee,Gerosa:2013laa,Gerosa:2017kvu,Fishbach:2017dwv,Vitale:2015tea,Stevenson:2017dlk,Talbot:2017yur,Farr:2017uvj,Farr:2017gtv,Mandel:2009nx,Belczynski:2017gds}. 




Besides spin directions, spin magnitudes carry important
information as well, since they depend on the angular momentum of the
BH's stellar progenitor and its evolution. At the moment, there
remains considerable uncertainty in BH spin measurements, with mild
tension between spins inferred from GW
observations~\cite{2016PhRvL.116f1102A,2016PhRvL.116x1103A,2017PhRvL.118v1101A,Abbott:2017oio,Abbott:2017gyy},
stellar evolution models~\cite{2015ApJ...810..101F} and X-ray binary
observations~\cite{Miller:2014aaa}. BH spins inferred from GW
observations to date have pointed towards slowly spinning BHs, while
inferences of BH spins from X-ray binaries tend to be higher,
including some inferred spins that are nearly
extremal~\cite{Gou:2013dna,McClintock:2006xd}, though these BHs need not be part of the same population~\cite{2012arXiv1208.2422B}. By nearly extremal, we
mean spins close to the theoretical maximum for a Kerr BH,
i.e., dimensionless spins $\chi$ satisfying
\begin{eqnarray}
\chi \equiv \frac{S}{M^2} \approx 1,
\end{eqnarray}
where $S$ is the spin angular momentum and $M$ is the mass of the spinning BH and throughout the paper we use units where $G=c=1$.

GW observations primarily provide information about the effective spin
$\chi_{\rm eff}$, a combination of the spin components along the
binary's orbital angular momentum that is conserved to second
post-Newtonian\footnote{The second post-Newtonian order is a term of
  order $(v/c)^4$ relative to the leading-order term, where $v$ is
  some characteristic velocity of the systems and $c$ is the speed of
  light.} order~\cite{Racine:2008qv,Gerosa:2015tea}. Specifically,
\begin{equation}
\chi_{\rm{eff}}=\frac{m_1 (\vec{\chi}_1 \cdot \hat{L}) + m_2 (\vec{\chi}_2 \cdot \hat{L}) }{m_1+m_2},
\end{equation} where $m_1$ and $m_2$ are the the masses of the larger and smaller BH respectively, $\hat{L}$ is a unit vector in the direction of the orbital angular momentum, and $\vec{\chi}_1$ and $\vec{\chi}_2$ are the dimensionless spin vectors of the BHs.
The apparent discrepancy between GW and X-ray binary measurements has
lead to stellar evolution models predicting a bimodality in the spin
distribution of BHs. These models suggest that some BHs in future LIGO
observations might have large spins that are also aligned with the
orbital angular
momentum~\cite{Zaldarriaga:2017qkw,Hotokezaka:2017esv}.

In this paper we pose the following question: if the BHs in a LIGO
source were to have nearly extremal spins, could we tell? To address
this question, we simulate GW signals using numerical-relativity (NR)
waveforms computed with the Spectral Einstein Code (SpEC)~\cite{spec-website}.
We use two SpEC simulations from the public Simulating eXtreme Spacetimes (SXS)
catalog~\cite{Mroue:2013PRL,SXSCatalog} and two new, previously unpublished simulations,
including one with the highest BH spins simulated to date. Three simulations have
 BH spin magnitudes nearly extremal and spin directions either both parallel to $\hat{L}$, both antiparallel to $\hat{L}$ or one spin parallel and one antiparallel.
 The fourth simulation has moderate BH spins and is included to help assess the impact of large
  individual spin magnitudes when they point in opposite directions.
We then use LIGO parameter estimation
methods and tools to infer the properties of the simulated signals,
including their masses and spins.

We find that current parameter estimation methods can recover large
spins, but only if the effective spin is large (meaning that the spins
are either both aligned or both antialigned with the orbital angular
momentum) and the signal-to-noise ratio (SNR) is sufficiently
high. The recovered spin magnitudes and effective spin are shifted significantly
towards less extremal values under the most commonly used spin prior
assumption. If, on the other hand, the effective spin is small
(meaning that the two BHs' spins point in opposite directions), we
accurately recover the small effective spin but incorrectly recover small individual
spins. Our results suggest that if the Universe contains BBH systems with nearly extremal spins, GW inference might fail to tell us.

The rest of this paper is organized as follows. In
Sec.~\ref{sec:signals}, we describe the NR waveforms, and the
simulated GW signals that we generate from them, as well as our
parameter estimation methods. In Sec.~\ref{sec:results}, we present
our results and discuss the conditions under which we can measure
large spins. We conclude in Sec.~\ref{sec:conclusion}.

\section{Methods}
\label{sec:signals}

We calculate our simulated gravitational waveforms using SpEC. SpEC's
methods, including recent improvements enabling more robust
simulations of merging BHs with nearly extremal spins, are described
in Ref.~\cite{Scheel2014} and the references therein.

We consider four numerical gravitational waveforms from
  merging BHs, each simulated with SpEC.  The BHs in
  each simulation have spins either aligned or
  antialigned with the orbital angular momentum. Two of these
  simulations (SXS:BBH:0305 and SXS:BBH:0306) were previously
  presented in
  Refs.~\cite{2016PhRvL.116f1102A,Lovelace:2016uwp,Bohe:2016gbl} and
  are available in the public SXS catalog~\cite{SXSCatalog}, while the
  other two (SXS:BBH:1124 and SXS:BBH:1137) are new.  The
configurations are summarized in Table~\ref{table:simulations}:
SXS:BBH:1124 has large spins aligned with the orbital angular
momentum; SXS:BBH:1137 has large spins antialigned with the orbital
angular momentum; SXS:BBH:0306 has two large spin pointing in opposite
directions, resulting in a small effective spin; and SXS:BBH:0305 has
moderate antiparallel spins and a small effective spin.

\begin{table}
\begin{centering}
\begin{tabular}{cccccccccccccccc}
\hline
\hline
\noalign{\smallskip}
SXS:BBH:  && $1/q$ &&  $\chi_{1z}$ &&  $\chi_{2z}$ &&$\chi_{\rm{eff}}$ && $N_{\rm orbits}$ && $f_{\rm{GW}}$(Hz) \\
\hline
\noalign{\smallskip}
0305                && 1.22 && 0.330   && -0.439 && -0.016 && 15.2 && 16.8   \\
0306               && 1.3 && 0.961  && -0.899 && 0.152 && 12.6 &&  19.4\\
1124		 && 1    && 0.998     && 0.998 &&  0.998 && 25 && 14.2 \\
1137            && 1    && -0.969     && -0.969 && -0.969 && 12 && 14.9\\
\noalign{\smallskip}
\hline
\hline
\end{tabular}
\end{centering}
\caption{Properties of the SXS simulations used in this paper. The
  table shows the mass ratio $q$, spin $\chi_1$ of the larger BH, spin $\chi_2$ of the smaller BH, the resulting $\chi_{\rm{eff}}$, the number of orbits
  $N_{\rm orbits}$ in the simulation, and the initial
  GW frequency of the $(\ell=2,m=2)$ mode for a system with a total mass of $70M_{\odot}$.}
\label{table:simulations}
\end{table}

We use the numerical-relativity (NR) data to simulate GW
signals as observed by the two Advanced LIGO
detectors with the projected sensitivity for the second observing
run~\cite{Aasi:2013wya}. As is common practice, we do not add detector noise on the simulated signal, which is equivalent to averaging over noise realizations~\cite{2010ApJ...725..496N}. All intrinsic parameters of the simulated
signals apart from the total mass are determined by the NR data and
are given in Table~\ref{table:simulations}. The total mass of the
system is an overall scale factor in vacuum general-relativity that we
are free to specify. We select extrinsic parameters such that the
orbital angular momentum of the binary points towards the GW
detectors\footnote{We have verified that this choice does not affect
  our results, since the signals we are studying are short and the
  effect of spin-precession is suppressed~\cite{Chatziioannou:2014bma,Chatziioannou:2014coa,Farr:2015lna}.} and place the binary systems over the Livingston
detector, scaling the source distance to achieve a signal-to-noise
ratio (SNR) of interest. See Ref.~\cite{Schmidt:2017btt} for a description of the details and implementation of the NR injection infrastructure we make use of.

We then analyze the simulated data with the parameter estimation
software library {\tt LALInference}~\cite{Veitch:2014wba}, which samples the
joint multidimensional posterior distribution of the binary
parameters. The posterior distribution is calculated through Bayes'
Theorem $p(\vec{x}|d)\sim p(\vec{x}) p(d|\vec{x})$, where
$p(\vec{x}|d)$ is the joint posterior for the parameters $\vec{x}$
given data $d$, $p(\vec{x})$ is the prior distribution, and
$p(d|\vec{x})$ is the likelihood for the data. In GW parameter
estimation and under the assumption of stationary and Gaussian
detector noise, the likelihood can be expressed as
$\text{ln} p(d|\vec{x}) \sim -1/2 (d-h(\vec{x})|d-h(\vec{x}))$, where
parentheses denote the noise-weighted inner
product~\cite{PhysRevD.49.2658} evaluated from a lower frequency of 20Hz (24Hz for SXS:BBH:0306) 
and $h(\vec{x})$ is the model for the GW signal.

\begin{figure}
\includegraphics[width=0.9\columnwidth,clip=true]{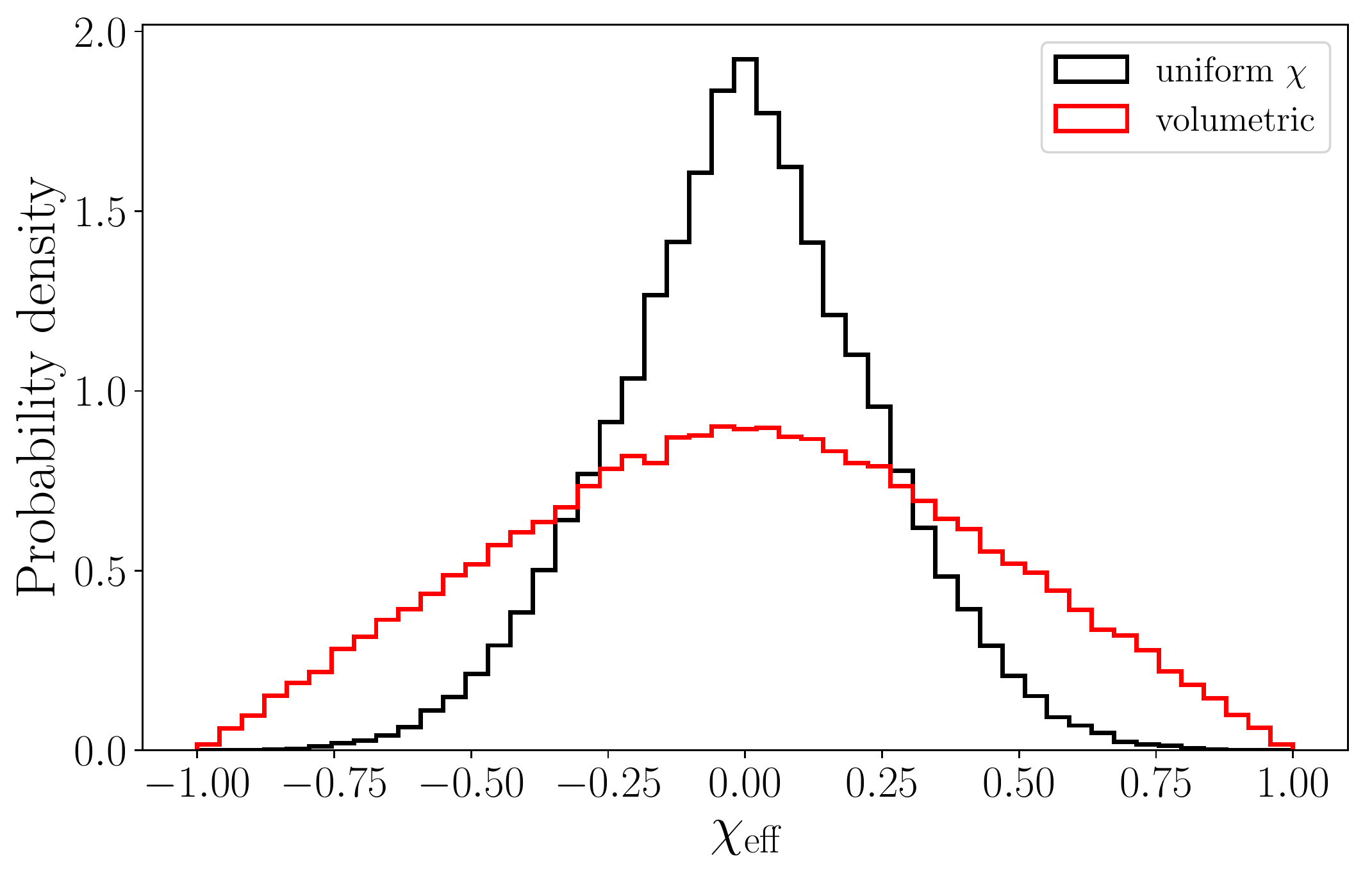}
\caption{ \label{fig:priors} Prior probability density for the
  effective spin when employing a uniform prior on spin magnitudes and
  directions (black, `uniform $\chi$'), and a uniform prior on spin components
  (red, `volumetric'). In both cases the prior on the component masses
  is flat.}
\end{figure}

The above procedure contains two important ingredients: the prior
distribution for the parameters $p(\vec{x})$ and a waveform model for
the GW signal $h(\vec{x})$. For the prior, we select a uniform
distribution for the sky location and the orientation of the source, a
uniform-in-volume distribution for the distance, and a uniform
distribution for the component masses. We explore two prior
distributions for the spin angular momenta. The first (a `uniform $\chi$'
prior) assumes that the spin magnitude and directions have a uniform
distribution, $p(\chi) d\chi \propto d\chi$ (this is the default choice for most GW analyses). The
second (a `volumetric' prior) assumes that the individual spin
components are uniformly distributed, $p(\chi) d\chi \propto \chi^2 d\chi$. The resulting prior distribution
for the effective spin from these two choices is plotted in
Fig.~\ref{fig:priors}. Both priors favor small
effective spins, though the volumetric prior has more support at high
$\chi_{\rm{eff}}$.  These prior choices affect parameter
inference~\cite{Vitale:2017cfs,Gerosa:2017mwk}, and we discuss their
impact on measuring large spins in Sec.~\ref{sec:results}.

We employ two waveform models in the {\tt LALInference} analysis, {\tt
  IMRPhenomPv2}~\cite{Hannam:2013oca} and {\tt
  SEOBNRv4}~\cite{Bohe:2016gbl}, which both include the inspiral, merger and ringdown phases of a
BBH coalescence. Both models have been extensively used for the analysis of GW
signals, see for example
Refs.~\cite{2017PhRvL.118v1101A,Abbott:2017oio}. {\tt IMRPhenomPv2}
includes the effects of spin-precession in an \emph{effective} way by
parameterizing it through a single effective parameter
$\chi_p$~\cite{Schmidt:2014iyl}. {\tt SEOBNRv4}, on the other hand,
assumes that the spins remain aligned with the orbital angular
momentum throughout the binary evolution. Both models have been calibrated against nonprecessing NR simulations (including a simulation with both spins at $0.994$ in the case of {\tt SEOBNRv4}) and have been shown
to match well the predictions of NR~\cite{Bohe:2016gbl}. Neither model
results in systematic biases in the case of
GW150914~\cite{Abbott:2016wiq,Bustillo:2016gid}. We choose to work with both waveform models both for computational convenience and as an independent cross-check of our results.

\section{Results}
\label{sec:results}

In this section, we present the results of the {\tt LALInference}
parameter estimation study performed on the simulated signals
described in Sec.~\ref{sec:signals}, and we discuss our
  ability to robustly characterize nearly extremal BHs in
  GW observations. Our results indicate that a standard parameter estimation study, such as the one employed by the LIGO and VIRGO collaborations, can lead to a reasonable estimation of the total mass, mass ratio, and effective spin of nearly-extremal BHs. However, we recover a systematic offset in $\chi_{\rm{eff}}$ away from extremality, which is compensated by a systematic bias in the total mass, an outcome of the mass--spin degeneracy. Our parameter estimates are summarized in Table~\ref{table:CIs} for both priors of Fig.~\ref{fig:priors}.

\begin{table}
\begin{centering}
\begin{tabular}{ccccc}
\hline
\noalign{\smallskip}
 & &  & \multicolumn{2}{c}{Recovered } \\
SXS:BBH: & Parameter & Injected  & `uniform $\chi$'  & `volumetric'   \\
\hline
\hline
\noalign{\smallskip}
0305  & \specialcell{$q$ \\ $M (\textrm{M}_{\odot})$ \\ $\chi_{\text{eff}}$ \\ $\chi_{1z}$ \\ $\chi_{2z}$} & \specialcell{0.82 \\ 70 \\ -0.016 \\ 0.330 \\ -0.439} & \specialcell{(0.65,1) \\${70.1}^{+2.4}_{-2.3}$\\${-0.013}^{+0.084}_{-0.099}$ \\${-0.002}^{+0.271}_{-0.258}$\\${-0.011}^{+0.274}_{-0.386}$ } & \specialcell{(0.65,1) \\${70.0}^{+2.6}_{-2.6}$\\${-0.007}^{+0.086}_{-0.094}$ \\ ${0.158}^{+0.551}_{-0.909}$ \\${-0.192}^{+0.989}_{-0.728}$ } 
 \\ \hline
0306 & \specialcell{$q$ \\ $M (\textrm{M}_{\odot})$ \\ $\chi_{\text{eff}}$ \\ $\chi_{1z}$ \\ $\chi_{2z}$} & \specialcell{0.77 \\ 70 \\ 0.152 \\ 0.961 \\ -0.899 } & \specialcell{(0.65,1) \\${69.5}^{+3.2}_{-2.9}$\\${0.176}^{+0.100}_{-0.103}$ \\${0.201}^{+0.388}_{-0.320}$\\${0.109}^{+0.446}_{-0.411}$ } & \specialcell{(0.63,1) \\${69.5}^{+2.8}_{-2.6}$\\${0.171}^{+0.089}_{-0.088}$ \\ ${0.484}^{+0.427}_{-0.820}$ \\${-0.209}^{+1.003}_{-0.701}$ } 
\\ \hline
1124 & \specialcell{$q$ \\ $M (\textrm{M}_{\odot})$ \\ $\chi_{\text{eff}}$ \\ $\chi_{1z}$ \\ $\chi_{2z}$} & \specialcell{1 \\ 70 \\ 0.998 \\ 0.998 \\ 0.998} &\specialcell{(0.63,1) \\${69.4}^{+1.9}_{-1.3}$\\${0.931}^{+0.035}_{-0.046}$ \\(0.89,1)\\(0.84,1) } & \specialcell{(0.61,1) \\${70.8}^{+1.9}_{-1.3}$\\${0.960}^{+0.029}_{-0.055}$ \\(0.94,1)\\(0.84,1) } 
\\ \hline
1137 & \specialcell{$q$ \\ $M (\textrm{M}_{\odot})$ \\ $\chi_{\text{eff}}$ \\ $\chi_{1z}$ \\ $\chi_{2z}$} & \specialcell{1 \\ 70 \\ -0.969 \\ -0.969 \\ -0.969} & \specialcell{(0.72,1) \\${73.8}^{+3.6}_{-2.7}$\\${-0.811}^{+0.150}_{-0.106}$ \\(-1,-0.65)\\(-1,-0.59) } & \specialcell{(0.77,1) \\${74.1}^{+2.9}_{-3.1}$\\${-0.790}^{+0.112}_{-0.144}$ \\(-1,-0.57)\\(-1,-0.60)  }  
\\ \hline
\noalign{\smallskip}
\end{tabular}
\end{centering}
\caption{Injected and recovered parameters for the four SXS simulations we study. For each simulated signal (first column) we quote the injected value (third column) and the recovered values (fourth and fifth column) for the mass ratio, the total mass, the effective spin, and the two spin components along the orbital angular momentum (third column). The fourth column shows results obtained with {\tt IMRPhenomPv2} and the `uniform $\chi$' prior, while the fifth column presents results with {\tt SEOBNRv4} and the `volumetric' prior. The recovered values we quote are either median and 90\% credible intervals or one-sided 90\% credible intervals, depending on whether the corresponding posterior rails agains a prior boundary, as further explained in, for example,~\cite{TheLIGOScientific:2018pe}.}
\label{table:CIs}
\end{table}
%

\subsection{Source characterization}

\begin{figure}
\includegraphics[width=0.9\columnwidth,clip=true]{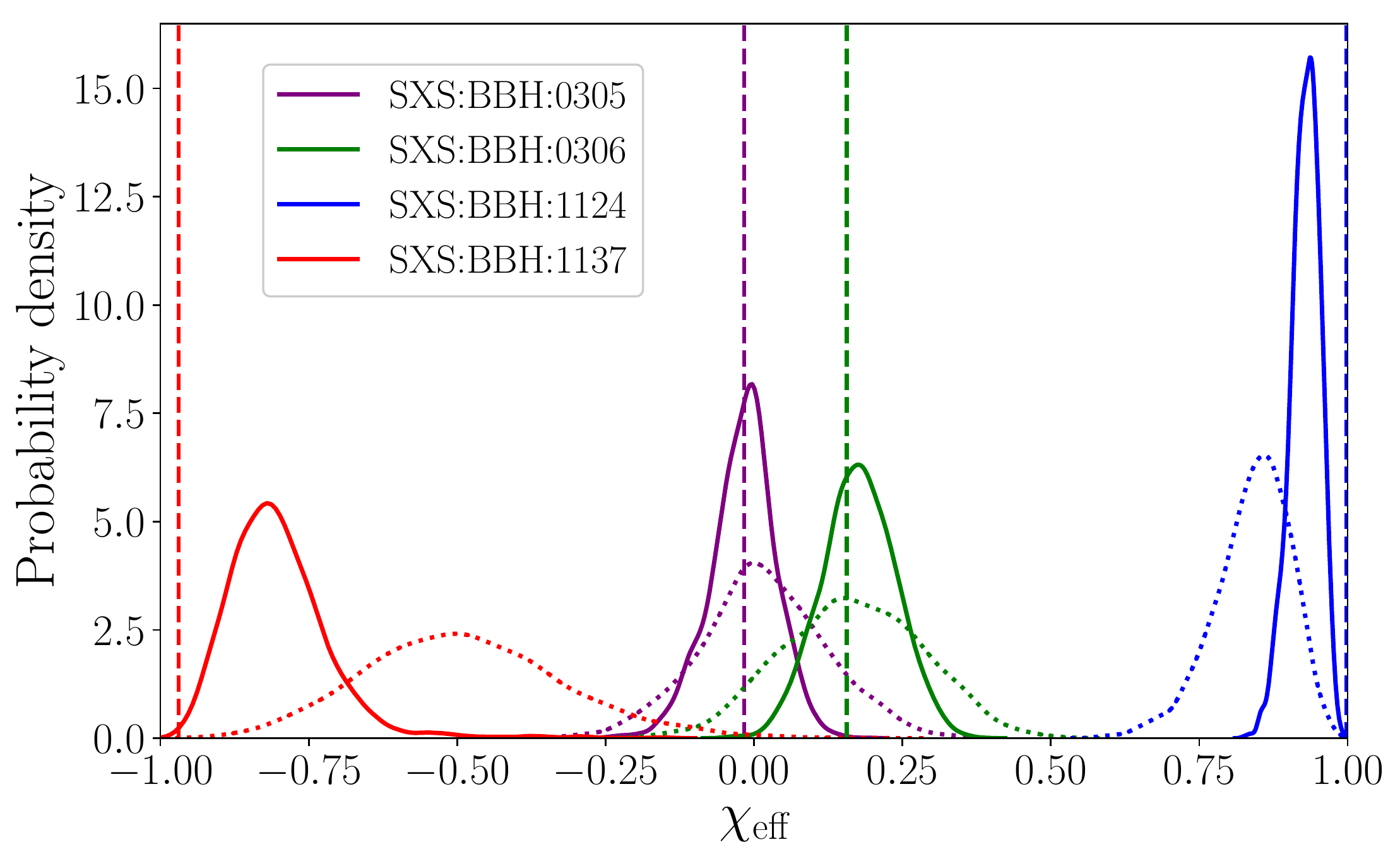}
\caption{ \label{fig:chieff_all_IMR} Marginalized posterior
  probability density for the effective spin for simulated
  signals at an
 SNR of $25$ (solid lines) and 12 (dotted lines),
  and a total mass of $70M_{\odot}$. The data are analyzed with {\tt
    IMRPhenomPv2} and the `uniform $\chi$' prior of
  Fig.~\ref{fig:priors} (see Fig.~\ref{fig:chieff_all_IMRvsEOBvsprior} for a reanalysis with the `volumetric' prior). The vertical dashed lines denote the true
  values of the effective spin. In all cases, the effective spin is
  measured, though this measurement is biased when the true value of
  $\chi_{\rm{eff}}$ is close to $\pm1$. For small values of the true effective
  spin, the posterior becomes more narrow as the SNR of the signal
  increases. For large (absolute value) effective spins, on the other
  hand, the posterior both becomes more narrow and shifts towards the
  true value as the signal becomes stronger.}
\end{figure}

The effective spin $\chi_{\rm{eff}}$ is one of the best measured spin parameters with
GWs. Therefore, it is commonly employed
to characterize spin measurability and to study the formation channels of
BBHs. Figure~\ref{fig:chieff_all_IMR} shows the marginalized posterior
probability density for $\chi_{\rm{eff}}$ for four simulated signals
using the NR simulations of Table~\ref{table:simulations}, analyzed
with the spin-precessing model {\tt IMRPhenomPv2}. In all cases, the
effective spin posterior is significantly different than the employed `uniform $\chi$' prior (see Fig.~\ref{fig:priors}) indicating that the posteriors are data driven.
However, this measurement is not accurate in the case where the true
$\chi_{\rm{eff}}$ value is close to the edges of its prior
range. Specifically, for cases SXS:BBH:1124 and SXS:BBH:1137, the
true value is outside the 99\% posterior credible interval, consistent also with the findings of Ref.~\cite{Afle:2018slw}. In the
next section we discuss this bias and its dependence on the specific
form of the default, `uniform $\chi$'
spin prior employed here, which disfavors large
$\chi_{\rm{eff}}$ values.

\begin{figure}
\includegraphics[width=0.9\columnwidth,clip=true]{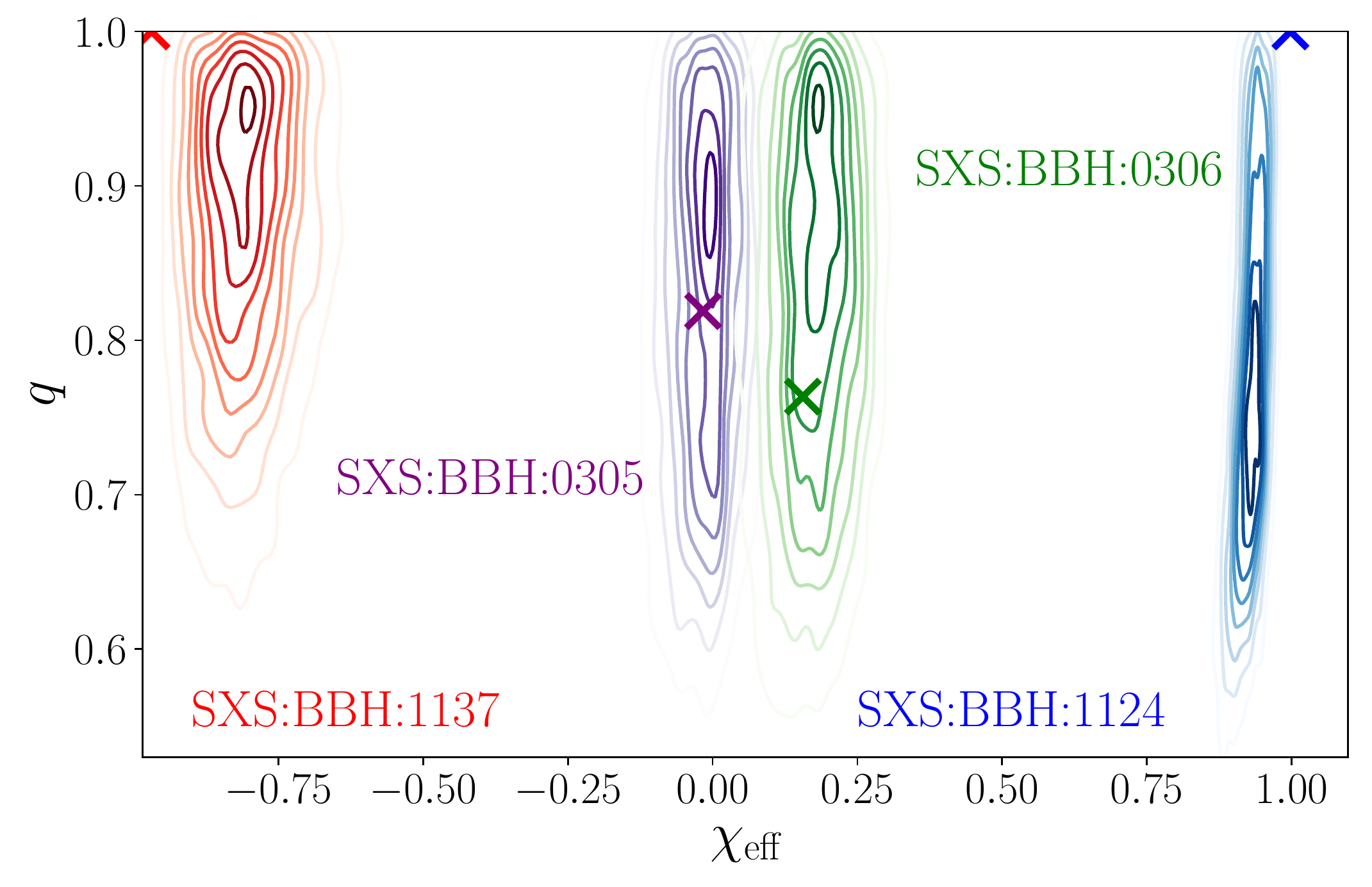}\\
\includegraphics[width=0.9\columnwidth,clip=true]{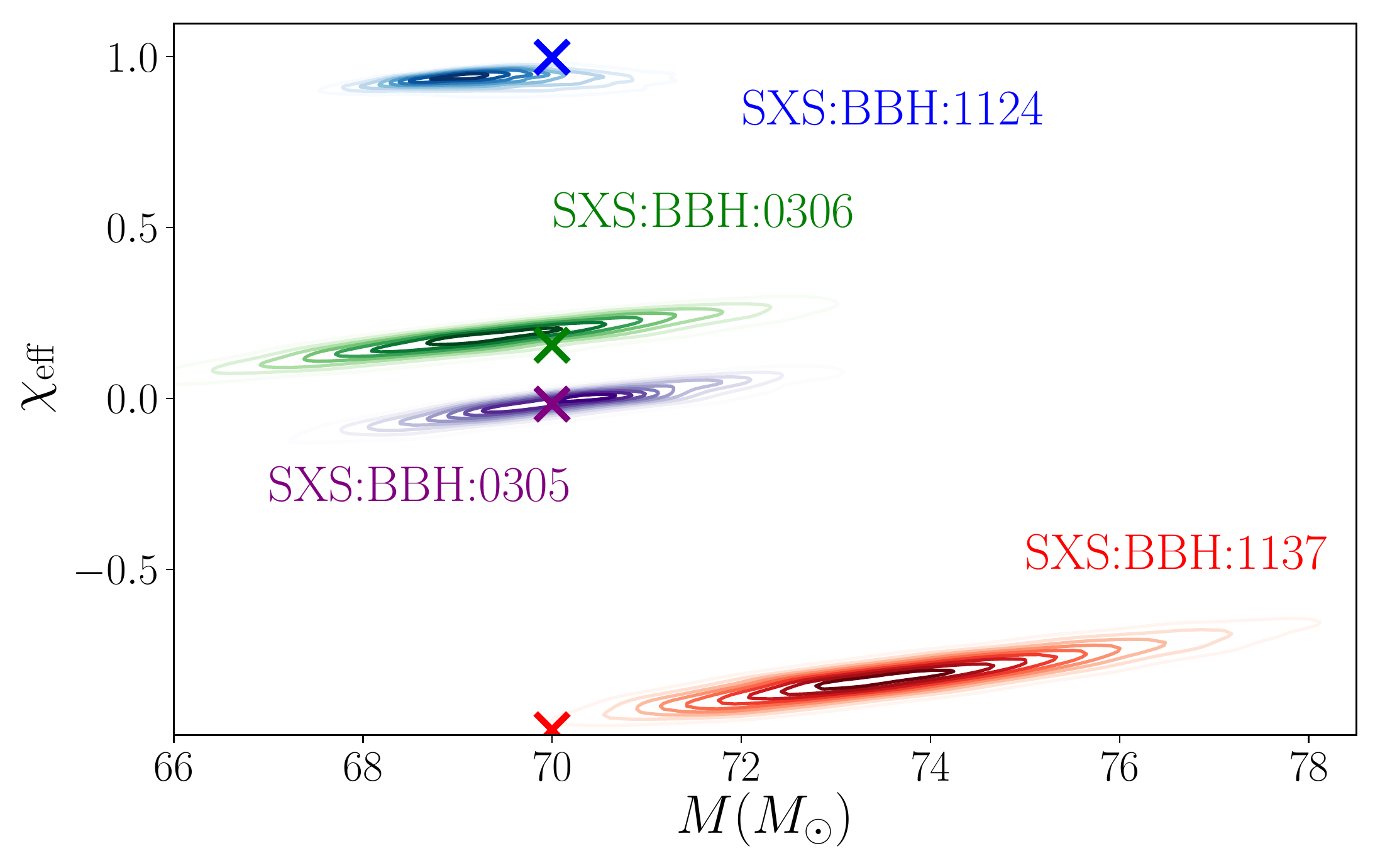}
\caption{ \label{fig:masses_all_IMR} Marginalized two-dimensional
  posterior probability density for the effective spin parameter and
  the mass ratio (top panel) and for the effective spin and the total
  mass (bottom panel) for four simulated signals at an SNR of $25$ and a total mass
  of $70M_{\odot}$.  The data are analyzed with {\tt IMRPhenomPv2} and
  the `uniform $\chi$' prior of Fig.~\ref{fig:priors}.  The true value is denoted
  with a cross of the same color as the corresponding contours.  For
  short signals such as the ones studied here, the effective spin is
  predominantly correlated with the total mass, as demonstrated in the
  bottom panel. This correlation is almost broken for the longest
  duration signal (SXS:BBH:1124, blue posterior) for which the
  effective spin shows a small correlation with the mass ratio (top
  panel).}
\end{figure}

Regarding the mass parameters, Fig.~\ref{fig:masses_all_IMR} shows the two-dimensional
posterior for the effective spin and the mass ratio (top panel), and
the total mass of the system (bottom panel)\footnote{In this and all similar two-dimensional plots with multiple level contours each line corresponds to a 10\% increment in the probability.}. It is well known that the
effective spin is correlated with either the mass ratio or the total
mass, depending on the duration of the signal~\cite{PhysRevD.49.2658}. For longer
signals that include a long inspiral phase, the effective spin is
correlated with the mass ratio, as they both affect the GW phase at
the same post-Newtonian order. On the other hand, if a signal consists
primarily of the merger phase, the effective spin is correlated with
the total mass, since they both affect the frequency of the
merger. This trend is visible in Fig.~\ref{fig:masses_all_IMR}, where
the $M-\chi_{\rm{eff}}$ correlation is more pronounced than the
$q-\chi_{\rm{eff}}$ one for all signals other than SXS:BBH:1124. Since SXS:BBH:1124 has a large positive spin angular momentum it is subject to the effect commonly called ``orbital hangup'', an
outcome of post-Newtonian spin-orbit
coupling~\cite{Damour01c,Kidder:1995zr} (cf. the discussion in
Sec.~4.2 of ~\cite{Scheel2014} and the references therein). This makes SXS:BBH:1124 last longer and be more inspiral-dominated, and hence more susceptible to the q--$\chi_{\rm{eff}}$ correlation.

\begin{figure}
\includegraphics[width=0.9\columnwidth,clip=true]{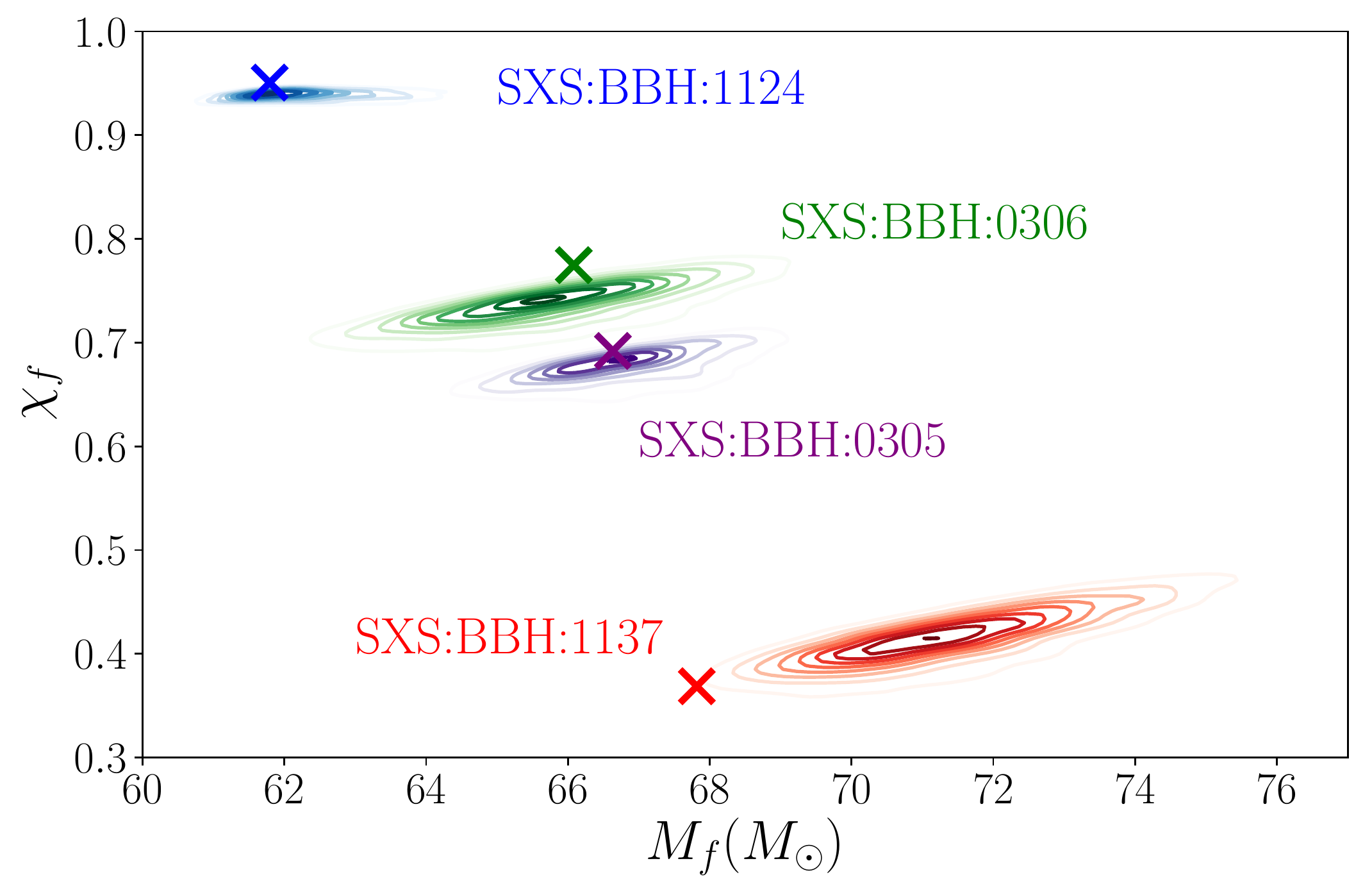}
\caption{ \label{fig:remnant_all_IMR} Marginalized two-dimensional
  posterior probability density for the mass and the spin of the remnant BH
  for four simulated signals at an SNR of $25$ and a total mass
  of $70M_{\odot}$.  The data are analyzed with {\tt IMRPhenomPv2} and
  the `uniform $\chi$' prior of Fig.~\ref{fig:priors}.  The true value is denoted
  with a cross of the same color as the corresponding contours. }
\end{figure}

Finally, the properties of the final remnant BH are examined in Fig.~\ref{fig:remnant_all_IMR} which shows the 
marginalized posterior distribution for the remnant mass and spin. As expected from the discussion of 
Fig.~\ref{fig:masses_all_IMR}, reliably extracting the properties of the final BH is challenging if the component spins
are large. Specifically, the true values are within the 90\% posterior credible region only in the SXS:BBH:0305 case.

%

\subsection{Spin measurability}

\begin{figure}
\includegraphics[width=\columnwidth,clip=true]{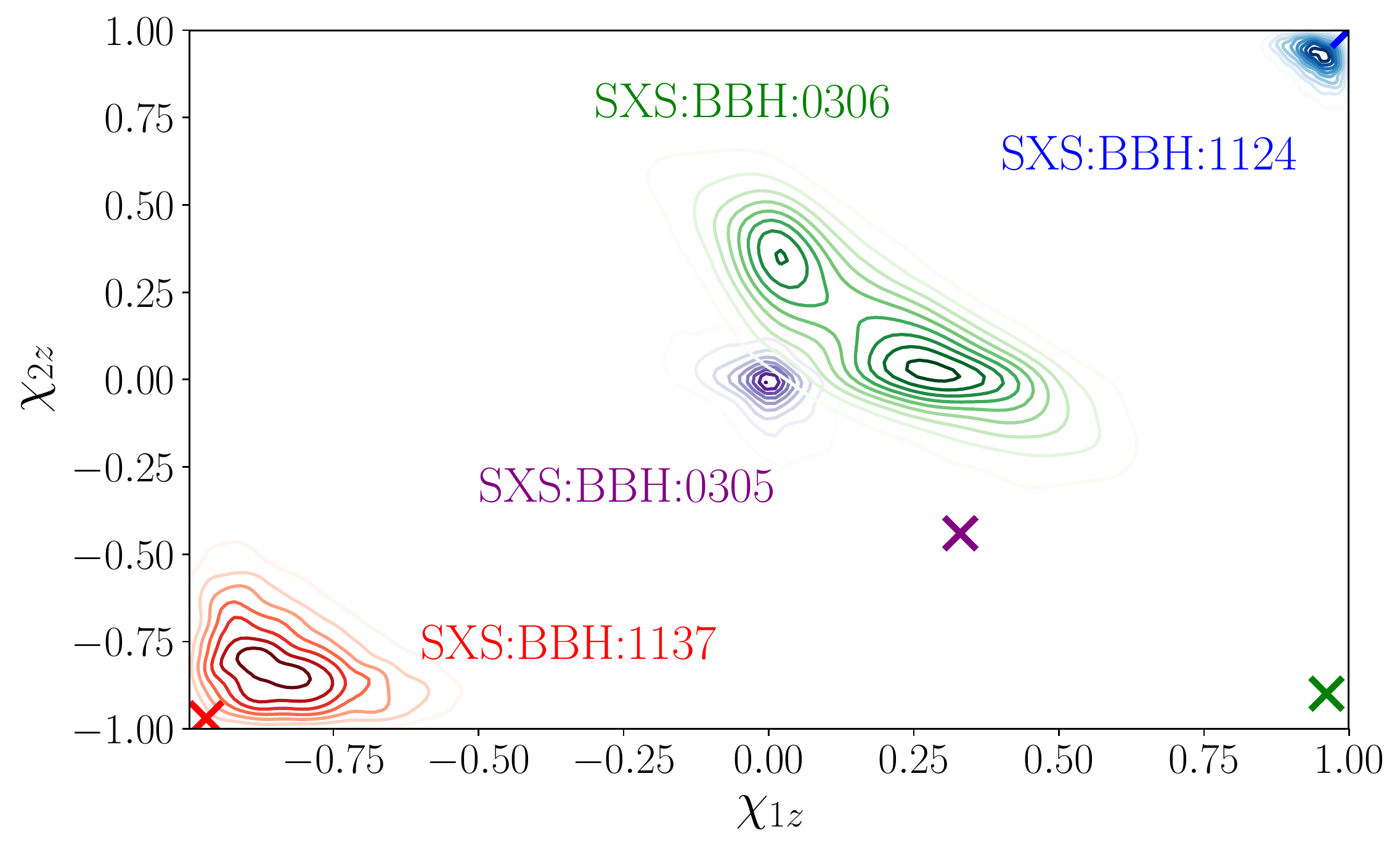}\\
\includegraphics[width=\columnwidth,clip=true]{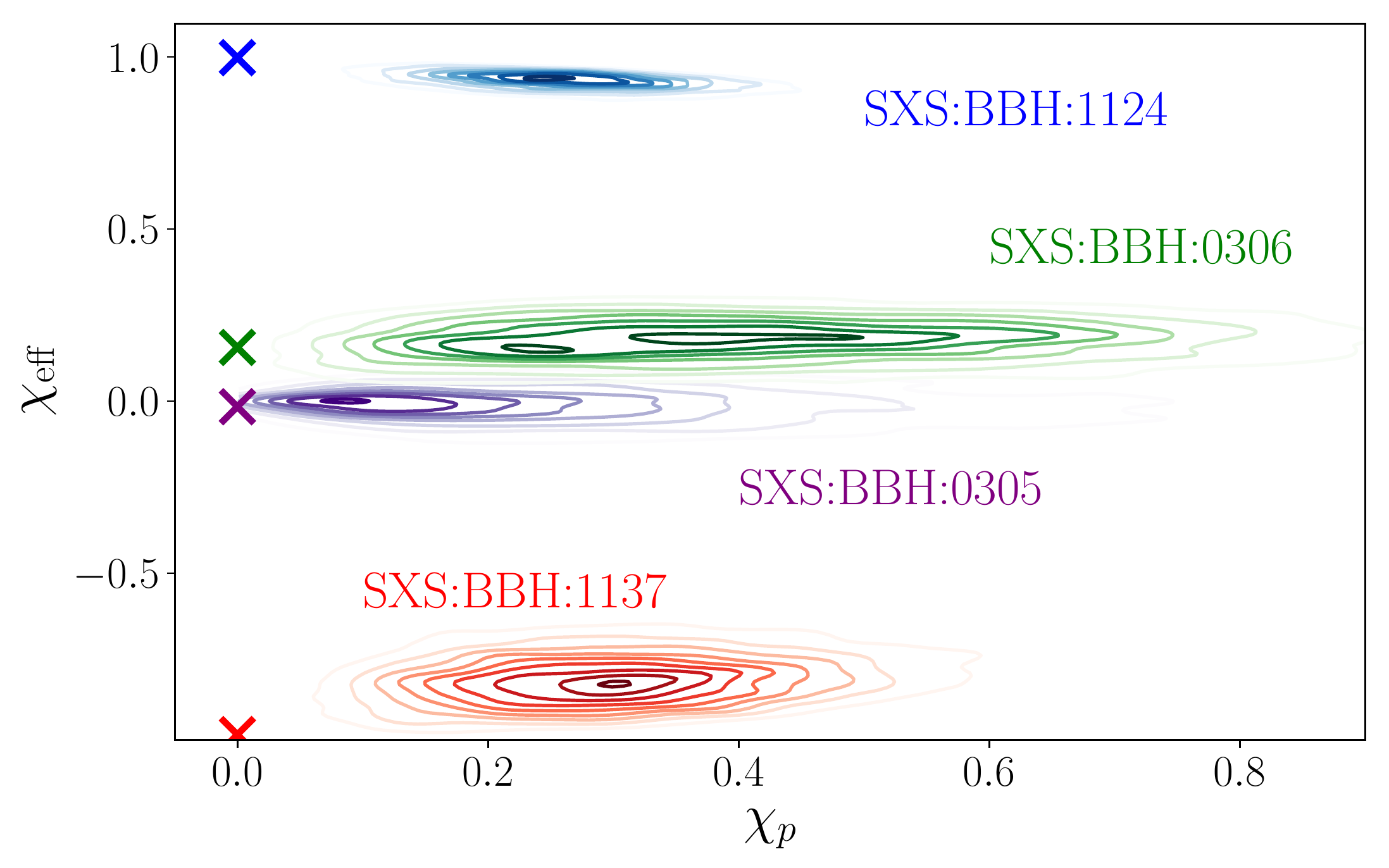}
\caption{ \label{fig:spins_all_IMR} Marginalized two-dimensional
  posterior probability density for the binary components' spins along
  the orbital angular momentum (top panel) and for the effective spin
  $\chi_{\rm eff}$ and $\chi_p$ (bottom panel). The data are analyzed with {\tt
    IMRPhenomPv2} and the uniform prior of Fig.~\ref{fig:priors}. The spin
  components are not recovered accurately, though the bias is less
  pronounced when the individual spins are large and both
  parallel or both antiparallel to the orbital angular momentum.}
\end{figure}

In the following we consider the measurability of various spin parameters in more detail. Figure~\ref{fig:spins_all_IMR} presents posterior probabilities for the
binary spin components along the orbital angular momentum (top panel)
and $\chi_{\rm{eff}}$ and $\chi_{p}$ (bottom panel).  Recall that the
spin parameter $\chi_{p}$ quantifies the amount of spin-precession
present in the system~\cite{Schmidt:2014iyl}. In each panel, we show
results for all four simulated signals at SNR 25; the true
parameters are shown as crosses in colors matching the corresponding
contours.  We find that the
large individual spin components can only robustly be measured when
both spins are large and parallel to each other. Conversely, configurations with spins antiparallel to each other are recovered as consistent with slowly-spinning binaries, as also alluded to by Fig.~\ref{fig:chieff_all_IMR}.
The
bottom panel of Fig.~\ref{fig:spins_all_IMR} shows that the $\chi_p$ posteriors extend to large values of $\chi_p$. However, comparison of these posteriors with the $\chi_p$ prior shows that the posterior is prior-dominated and we cannot constrain $\chi_p$ from the data~\cite{TheLIGOScientific:2016wfe}.

\begin{figure*}
\includegraphics[width=0.9\columnwidth,clip=true]{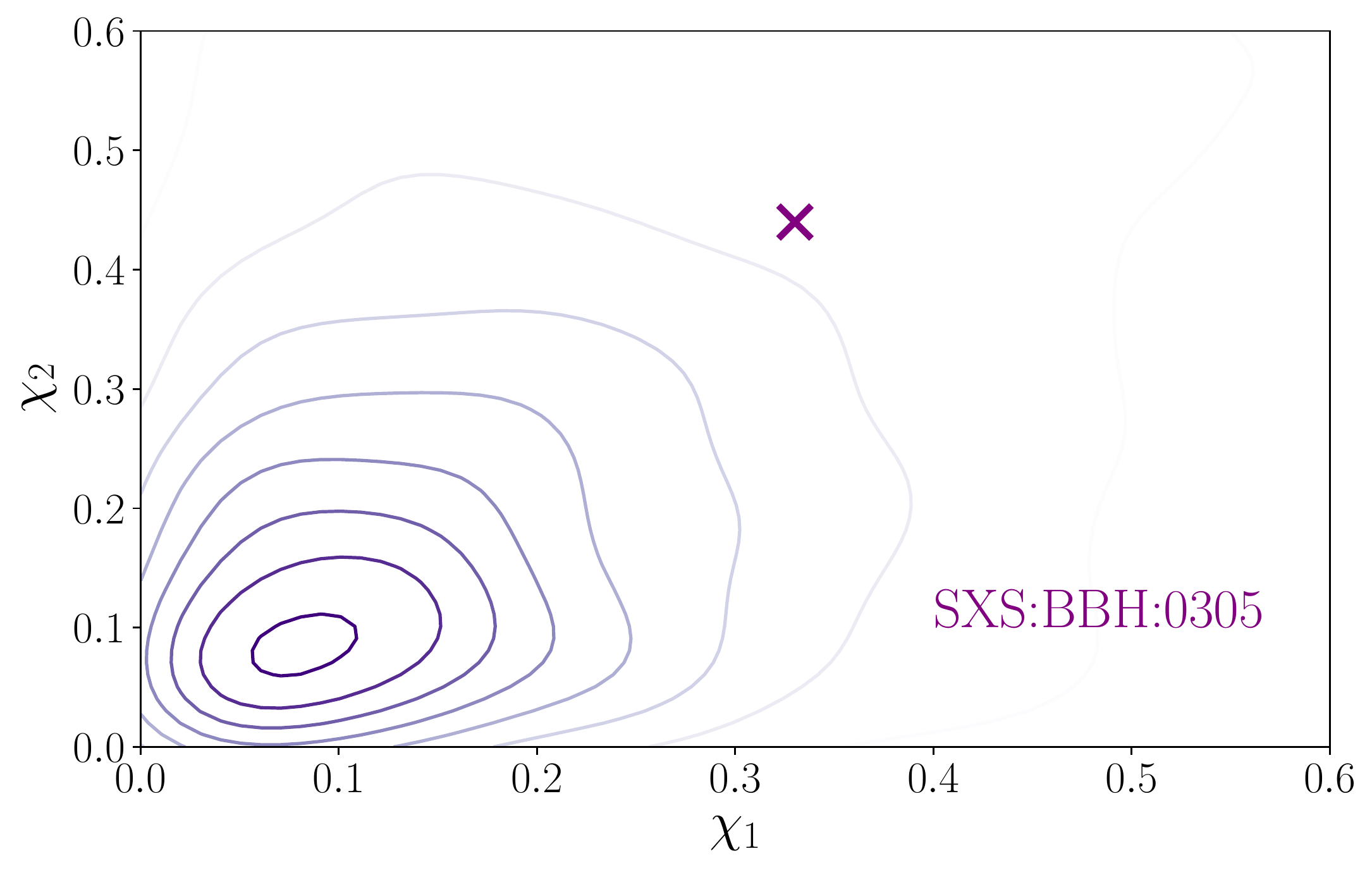}
\includegraphics[width=0.9\columnwidth,clip=true]{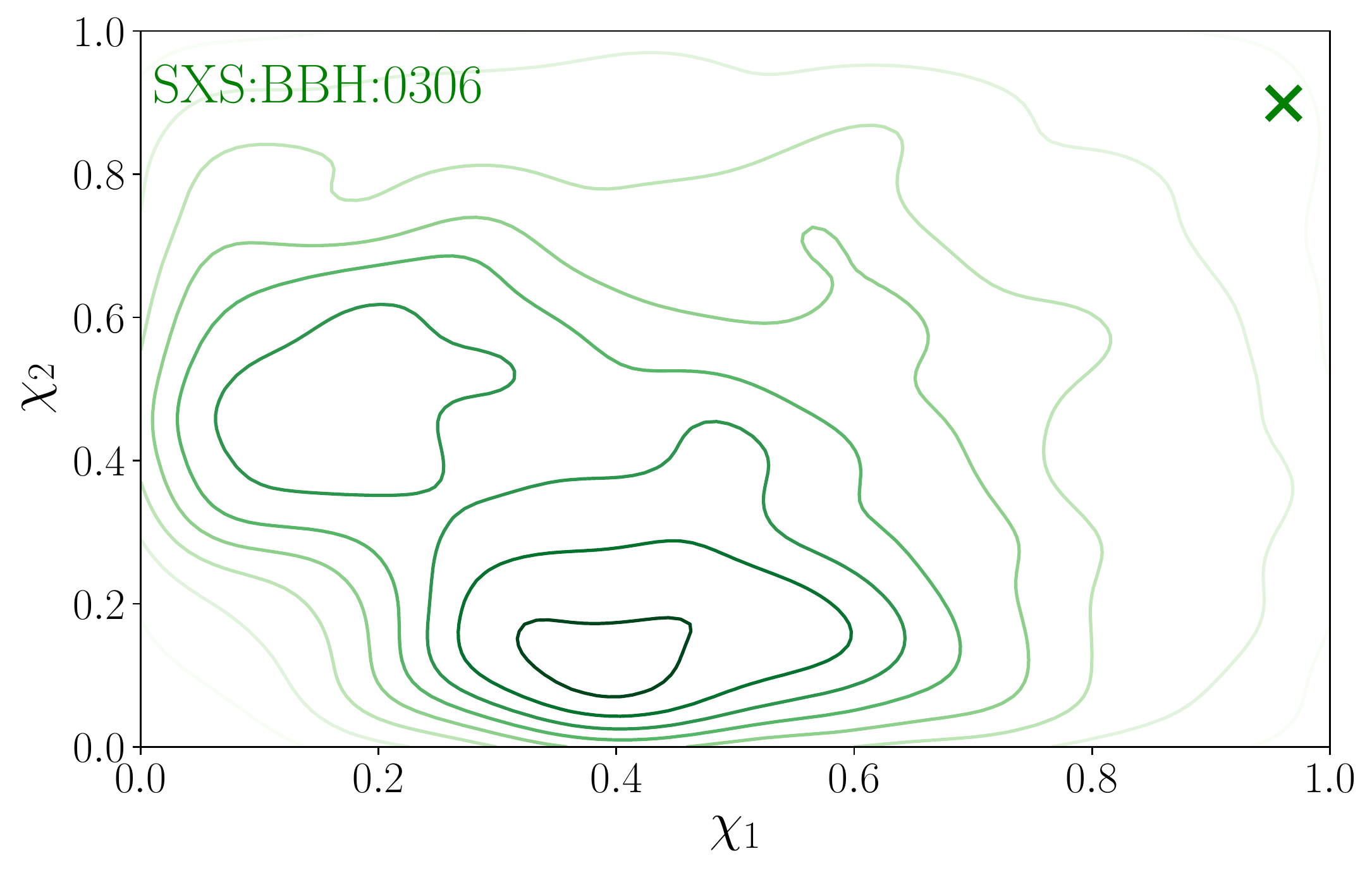}\\
\includegraphics[width=0.9\columnwidth,clip=true]{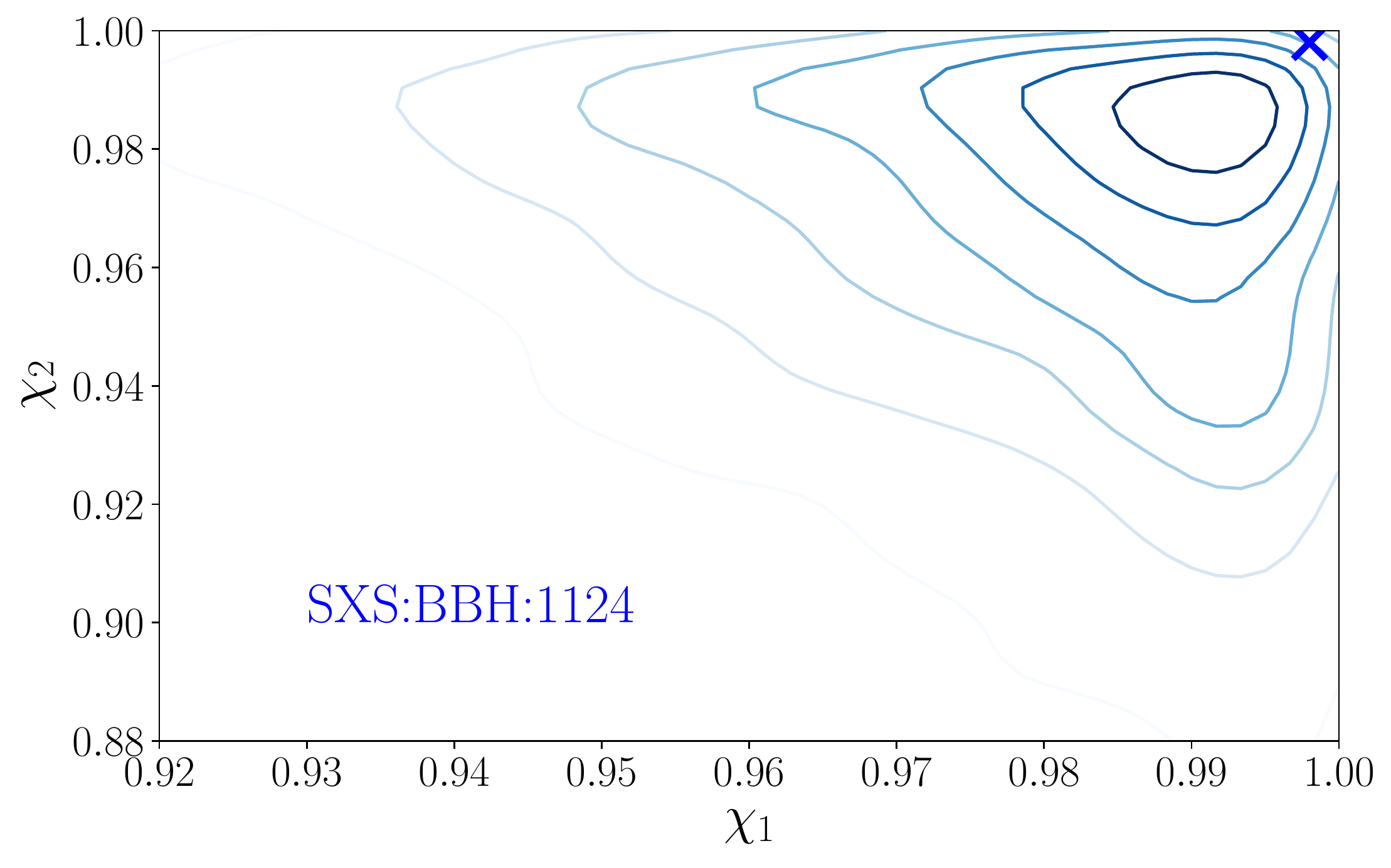}
\includegraphics[width=0.9\columnwidth,clip=true]{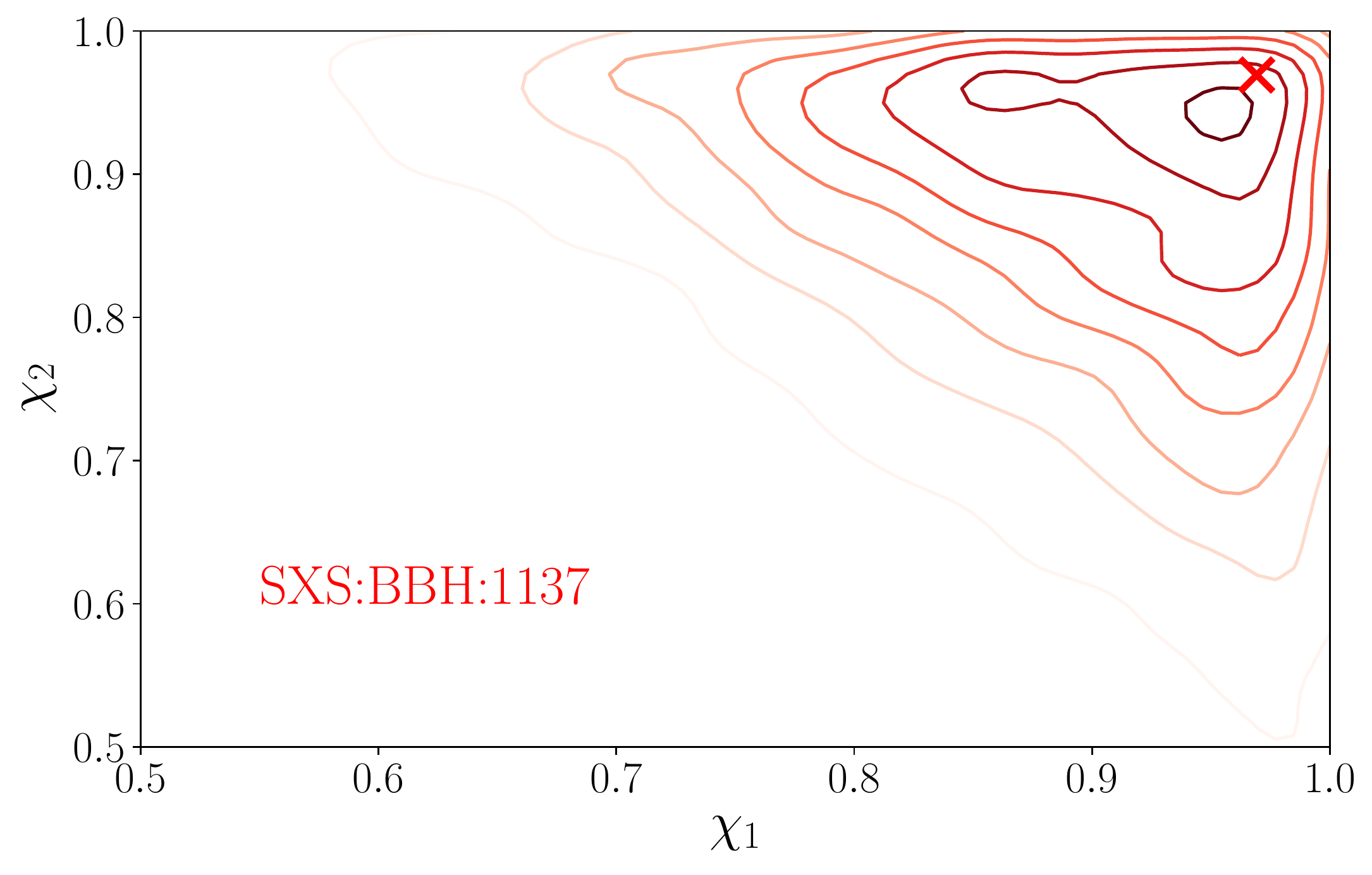}
\caption{ \label{fig:a1-a2_all_IMR} Marginalized two-dimensional
  posterior probability density for the binary components' spin
  magnitudes for four simulated signals with an SNR of $25$ and a
  total mass of $70M_{\odot}$. Here $\chi_1$ and $\chi_2$ are the
  spins of the larger and smaller BH, respectively.  The data
  is analyzed with {\tt IMRPhenomPv2} and the uniform prior
  of
  Fig.~\ref{fig:priors}. The true value is denoted with a
  cross symbol. Large individual spins can be measured
  when the spins are either both parallel or both antiparallel
  to the orbital angular momentum (1124 and 1137) but not
  when one spin is parallel and the other antiparallel (0305 and 0306).}
\end{figure*}

Figure~\ref{fig:a1-a2_all_IMR} examines the individual spin magnitudes and shows contours of the two-dimensional
posterior probability density for the individual spin magnitudes for
the four simulated signals with SNR 25.  Crosses indicate the true values
for the spins. The recovered individual spins are high when the true spins are
nearly extremal and parallel to each other but not when the true spins are
antiparallel to each other. This suggests that the individual spin magnitudes of
rapidly spinning BHs can only be reliably measured if the spins point
in the same direction, creating a larger effective spin $\chi_{\rm eff}$, in agreement with Fig.~\ref{fig:spins_all_IMR}.
The difficulty of measuring individual spin
magnitudes in general has been previously discussed in Ref.~\cite{Purrer:2015nkh}.

%

\subsection{Effect of spin prior}
\label{subsec:prior}

The accuracy of the measurement of the effective spin parameter in
Fig.~\ref{fig:chieff_all_IMR} is poor for the two cases where the 
true value is close to $\pm 1$. In this section, we discuss the
effect of the spin prior on the measurement of large spin
values~\cite{Vitale:2017cfs,Gerosa:2017mwk,Williamson:2017evr}.

Returning to Fig.~\ref{fig:chieff_all_IMR}, the dotted lines show the
marginalized posteriors for $\chi_{\rm{eff}}$ for signals of SNR
12. Stronger signals enable better parameter measurement and more
narrow posterior distributions. This expectation is confirmed for all
four systems studied here. Moreover, in the case of SXS:BBH:1124 and
SXS:BBH:1137 the posterior not only becomes more narrow, but it also shifts closer to the true value
demonstrating the difficulty of measuring large spins.

\begin{figure}
\includegraphics[width=\columnwidth,clip=true]{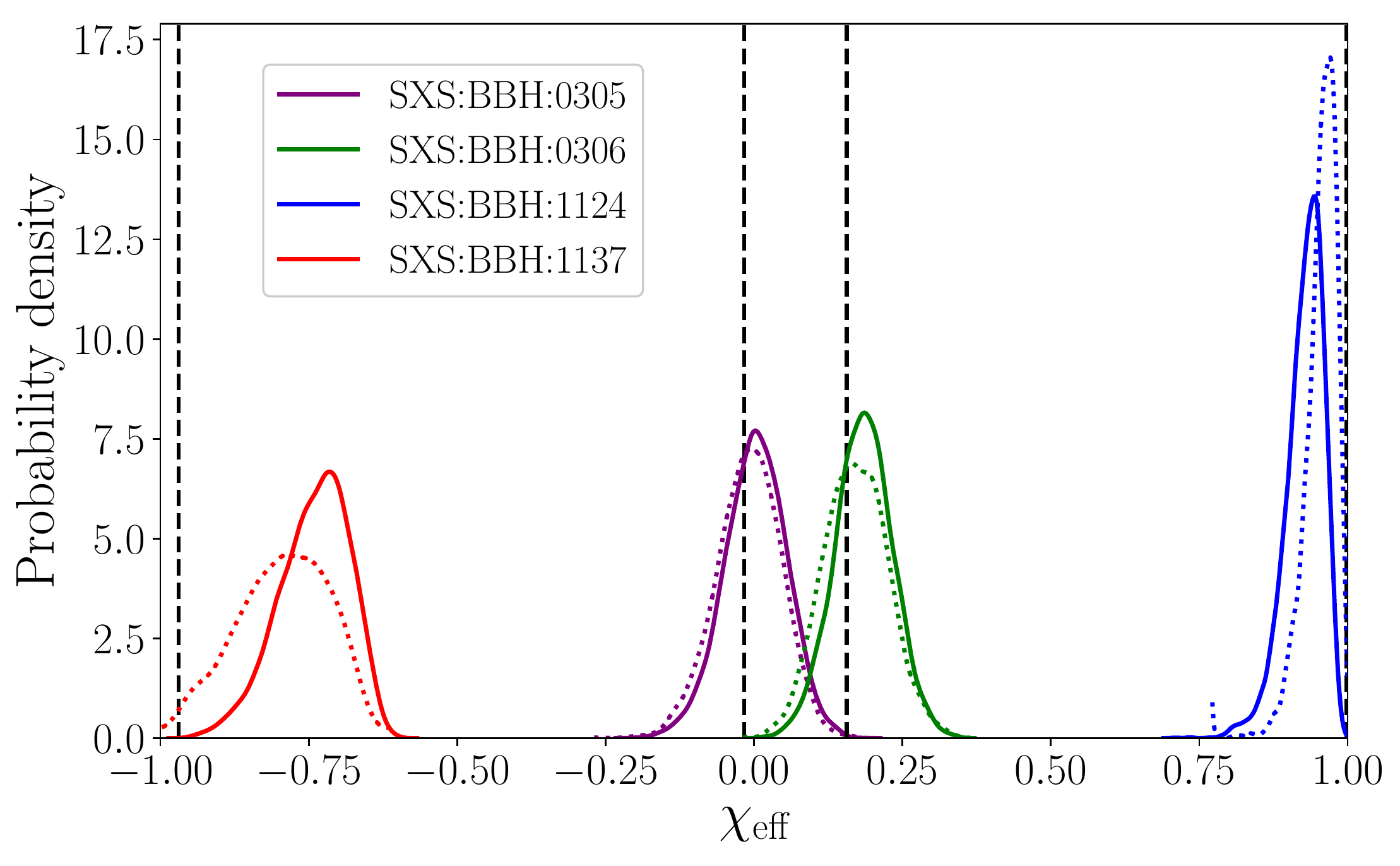}
\caption{ \label{fig:chieff_all_IMRvsEOBvsprior} 
Effective spin posteriors for two choices of the spin prior. Solid lines indicate the `uniform
  $\chi$' prior and dotted lines denote the `volumetric' prior. This analysis is performed with the spin-sligned waveform model {\tt
    SEOBNRv4}.}
\end{figure}

To explore the effect of prior we employ the `volumetric' prior
of Fig.~\ref{fig:priors}, which results in higher prior
probability at higher spins, as demonstrated in
Fig.~\ref{fig:priors}. Figure~\ref{fig:chieff_all_IMRvsEOBvsprior}
shows the posterior distribution for the effective spin for signals
analyzed with the spin-aligned waveform model {\tt SEOBNRv4} with the
`uniform $\chi$' (solid lines) and the `volumetric' (dotted lines) spin
prior\footnote{Despite the `uniform $\chi$' and `volumetric' priors being derived in the context of 3-dimensional spin vectors, we can still apply them to spin-aligned waveform models that only include a single spin degree of freedom, the spin component along the orbital angular momentum $\chi_{iz}$. In that case the prior on the sole spin degree of freedom is the same as the prior on the $\chi_{iz}$ spin component under the `uniform $\chi$' or `volumetric' priors.}.
In the SXS:BBH:0305 and SXS:BBH:0306 cases, all posteriors are
very similar, suggesting that the prior distribution has a lesser
effect on the posterior when the effective spin is small.

\begin{figure}
\includegraphics[width=0.9\columnwidth,clip=true]{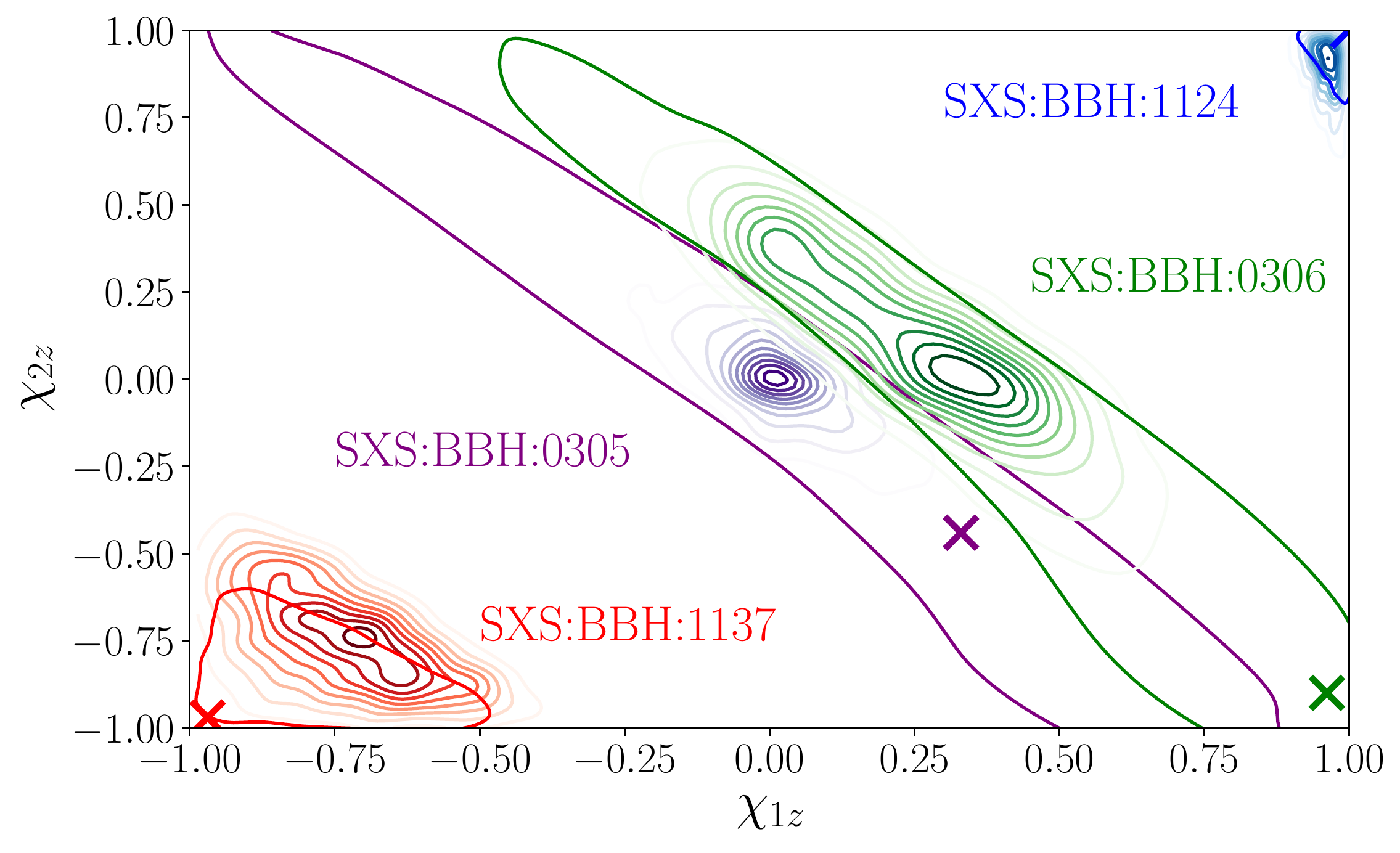}
\caption{ \label{fig:a1z-a2z_all_IMRvsEOBvsprior} Impact of spin prior on $\chi_{1z}-\chi_{2z}$ recovery with the aligned-spin waveform model {\tt
    SEOBNRv4}. Shown are incremental contours for the `uniform
  $\chi$' prior and a solid-line 90\% credible level contour for the `volumetric' prior. }
\end{figure}

In the case where $\chi_{\rm{eff}} \sim \pm 1$ on the other hand, the
choice of prior has a direct impact on the accuracy of the
measurement. Both for SXS:BBH:1124 and SXS:BBH:1137 the `volumetric'
prior leads to posteriors that have more support for larger
effective spin values, which are now within the 99\% posterior credible interval. A similar conclusion can be drawn from
Fig.~\ref{fig:a1z-a2z_all_IMRvsEOBvsprior}, which shows 
contours for the two-dimensional posterior probability density for the
spin components along the orbital angular momentum. As expected, all posteriors derived
with the `volumetric' prior have more support for large values of the
spin components.

%

\subsection{Effect of signal duration}

Due to the finite length of the NR data, all results presented in
the above subsections assumed a total mass of $70M_{\odot}$, which is
 comparable to the total mass of
GW150914~\cite{2016PhRvL.116f1102A}. If the total mass of the system
is lower than this value, the start of the numerical waveform falls within the
sensitive frequencies of the detector, potentially affecting the
results of parameter estimation~\cite{Mandel:2014tca}. To study the effect of the signal
duration on our results, instead, we use the waveform model {\tt
  IMRPhenomPv2} to simulate the GW data with parameters equal to those
of SXS:BBH:1124 and SXS:BBH:1137 but with a total mass of $30,50,$ and
$70 M_{\odot}$. We employ the same model for signal recovery and find
that the resulting posteriors are very similar, suggesting that our
main conclusions are unaffected by the signal duration.


\subsection{Model accuracy}

Our study suggests that current analyses are sub-optimal for
characterizing signals with large spins. However, the waveform models
used for these analyses may also lose accuracy at this
challenging region of the parameter space~\cite{Bohe:2016gbl}. This
prompts the question: is it hard to measure large spins because of the
posterior properties or because the waveform models employed
misbehave? In order to fully address this question we would have to perform parameter estimation 
directly using NR waveforms, something that is currently impossible for the region of the 
parameter space we are interested in. However, below we discuss evidence suggesting that the
difficulty to measure large spins has less to do with the accuracy of
the models, and more with the properties of the likelihood function and the prior choices.

First, when the SNR of the signal is increased, the posterior distribution for 
$\chi_{\rm{eff}}$ in the SXS:BBH:1137 and SXS:BBH:1124 cases shifts towards the true value, as
shown in Fig.~\ref{fig:chieff_all_IMR}. Since systematic errors caused by model inaccuracies 
do not depend on the SNR, the shift in the posterior suggests it is mainly the prior that 
keeps the posteriors away from large $\chi_{\rm{eff}}$ values.

\begin{figure}[h!]
\includegraphics[width=0.9\columnwidth,clip=true]{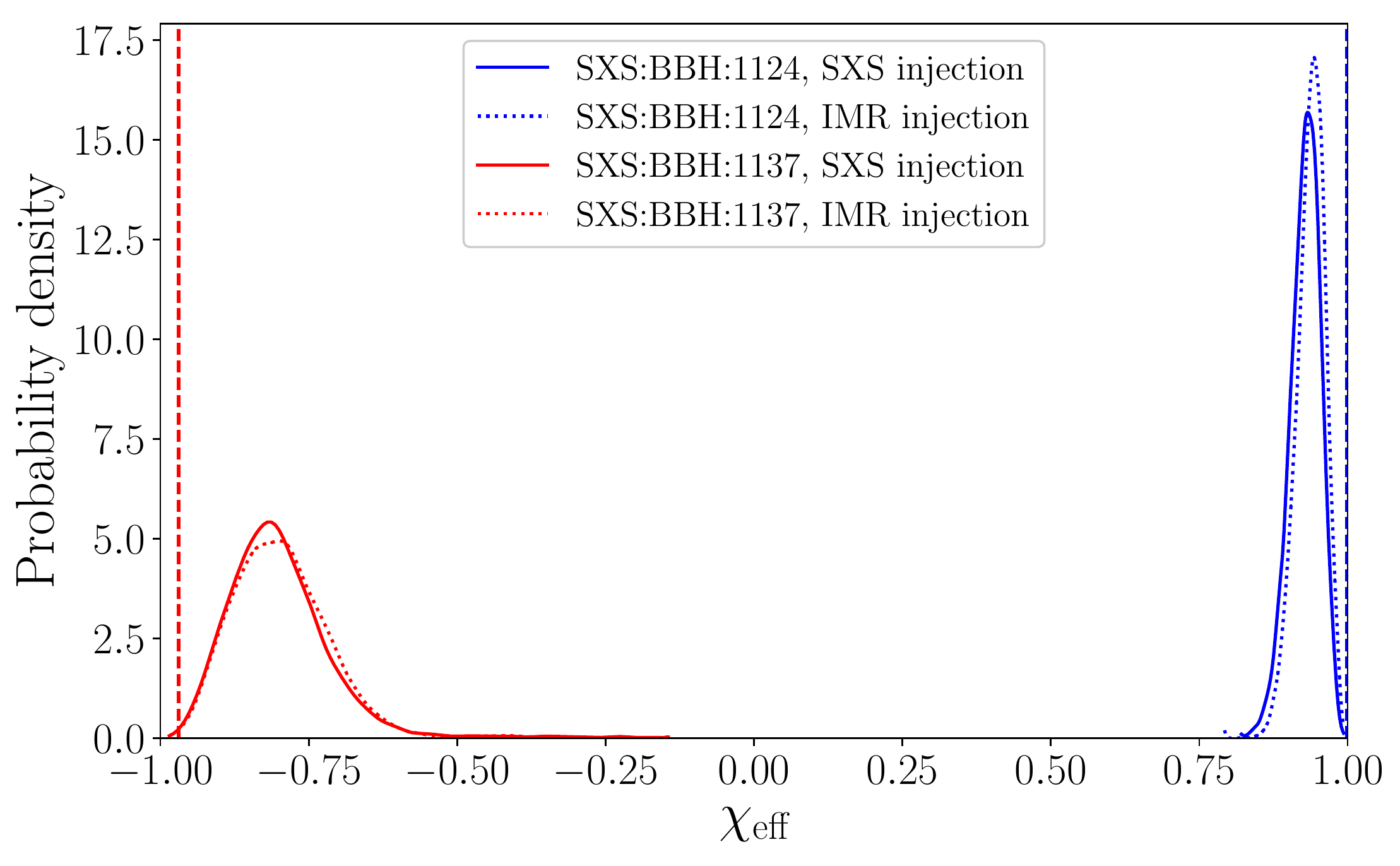}
\caption{ \label{fig:SXSvsIMR} Similar to Fig.~\ref{fig:chieff_all_IMR} for signals created with NR data (solid lines), and with the {\tt IMRPhenomPv2} waveform model (dotted lines). The similarity between the solid and the dotted curves suggest that systematic difference between NR and waveform models are not the dominant cause of our conclusion that large spins are difficult to measure.}
\end{figure}

Second, we repeat the analysis described above and compute the
posterior for the effective spin parameter using a simulated signal
created with the numerical waveforms and with the {\tt IMRPhenomPv2}
waveform model. We find a large similarity between the posterior
obtained with the different data, as shown in Fig.~\ref{fig:SXSvsIMR}. Specifically, the shift in the posterior 
due to the change of data in Fig.~\ref{fig:SXSvsIMR} is smaller than the shift due to changing the spin prior in Fig.~\ref{fig:chieff_all_IMRvsEOBvsprior}.
This suggests that at the injected parameter values the NR waveform and the data created with
{\tt IMRPhenomPv2} do not possess noticeable differences as far as parameter estimation 
is concerned.

\begin{figure}[h!]
\includegraphics[width=0.9\columnwidth,clip=true]{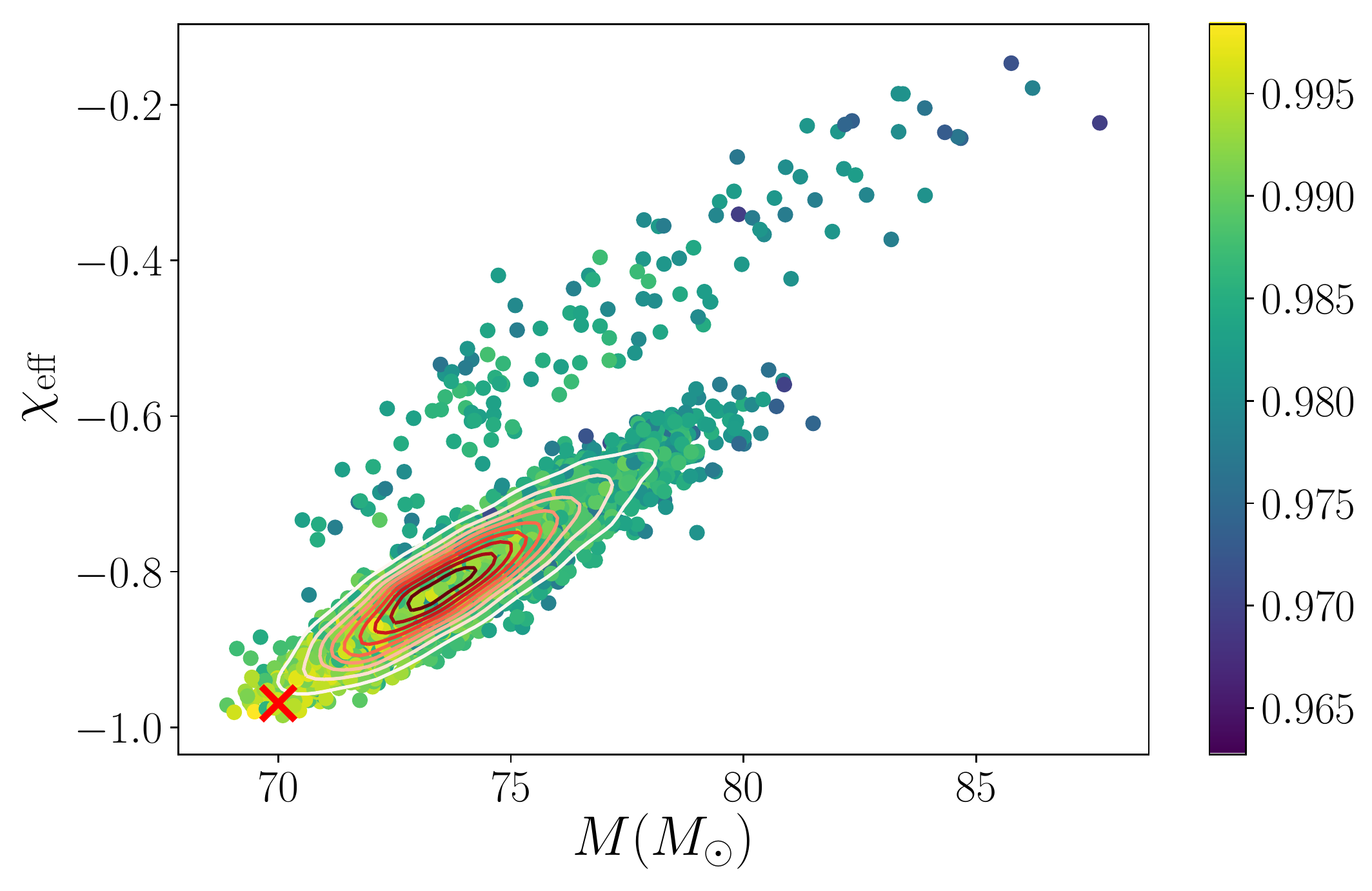}
\caption{ \label{fig:overlap} Similar to the bottom panel of Fig.~\ref{fig:masses_all_IMR} for the case of SXS:BBH:1137 with 5000 scattered posterior samples colored by the value of their overlap with the simulated data. The overlap achieved close to the injected value is in the 99.5\% range.}
\end{figure}

Third, we employ a figure of merit commonly used in waveform modeling, namely the overlap between the signal and the template, defined as $(d|h(\vec{x}))/\sqrt{(d|d)(h(\vec{x})|h(\vec{x}))}$. Figure~\ref{fig:overlap} shows a scatter plot of the posterior samples for the SXS:BBH:1137 case of the lower panel of Fig.~\ref{fig:masses_all_IMR}. The samples are colored by their overlap value; we find overlaps around $99.5\%$ in the region of the injected parameters, and they drop as we move away from the true parameters. We obtain similar results for the other three NR signals studied here and the {\tt SEOBNRv4} model. The high value of overlap further suggests that systematic biases are subdominant for this region of the parameter space and for this SNR value~\cite{Chatziioannou:2017tdw}.

%

\section{Conclusions}\label{sec:conclusion}

In this paper, we assess the prospects of extracting the spins
of nearly extremal BHs in binaries with GW measurements. We find that
measurement of large spins is challenging. Favorable conditions
occur when both spins are large and parallel to each other, but even in
this case our posteriors are biased away from extremal effective spins. 
We argue that this is due to the commonly used spin
priors that disfavor large spins. 

Additionally, extremal spins are close to the edge of the spin
priors. This situation is similar to the case of measuring the mass ratio (or the symmetric mass ratio) of equal-mass systems. In fact, when the posterior distribution 
of a parameter rails agains a prior edge, it is customary to use one-sided credible intervals
or highest-probability-density intervals (for an extended discussion, see~\cite{TheLIGOScientific:2018pe}).
 However, we find this is not the case for the effective spin, 
since its posterior typically does not rail against the prior edge (see Fig.~\ref{fig:chieff_all_IMR}). 
We attribute this to the spin prior, which drops to vanishingly small values as $\chi_{\rm eff} \to \pm 1$.
In order to overcome this trend and obtain a likelihood-dominated effective spin posterior, 
a signal with large SNR is needed. 

Our results
showcase again the importance of priors and prior bounds in GW inference and suggest
the use of a wide range of spin priors. This will not only allow us to
study physical effects such as the large spins described here, but can
also enable further studies such as the hierarchical analysis
described in~\cite{Farr:2017uvj}.

\section{Acknowledgments}

We are pleased to thank Sebastian Khan and Jacob Lange for useful
discussions on producing simulated GW signals with
NR data. We would also like to thank Joshua Smith
and Jocelyn Read for helpful discussions and Leo Stein and Juan Calderon Bustillo for comments on the manuscript. This work was supported in
part by National Science Foundation grants PHY-1606522 and PHY-1654359
to Cal State Fullerton. We gratefully acknowledge
support for this research at Caltech from NSF grants PHY-1404569, PHY-1708212, and PHY-1708213 and the Sherman Fairchild Foundation
and at Cornell from NSF Grant PHY-1606654 and the Sherman
Fairchild Foundation.

\bibliography{Refs}

\end{document}